\documentclass{emulateapj}
\usepackage{apjfonts}

\newcommand{\etal}{et al.~}
\def\gsim{\lower 2pt \hbox{$\, \buildrel {\scriptstyle >}\over
{\scriptstyle \sim}\,$}}
\def\lsim{\lower 2pt \hbox{$\, \buildrel {\scriptstyle <}\over
{\scriptstyle \sim}\,$}}

\def\fuse{{\sl FUSE}}

\def\chandra{{\sl Chandra~}}

\def\nvi{N~{\scriptsize VI}}
\def\nvii{N~{\scriptsize VII}}

\def\oviii{O~{\scriptsize VIII}}
\def\ovii{O~{\scriptsize VII}}
\def\ovi{O~{\scriptsize VI}}
\def\oi{O~{\scriptsize I}}

\def\neix{Ne~{\scriptsize IX}}
\def\nex{Ne~{\scriptsize X}}

\def\fexvii{Fe~{\scriptsize XVII}}

\newcommand{\persqrcm}{{\rm cm}^{-2}}

\def\xs{LMC~X--3}

\shortauthors{Yao \etal}
\shorttitle{X-ray and UV spectroscopy of Galatic diffuse hot gas}

\begin{document}
\slugcomment{\em Accepted for publication in the  Astrophysical Journal}

\title{X-ray and UV spectroscopy of Galactic diffuse hot gas along the 
LMC~X--3 sight line}
\author{Y. Yao\altaffilmark{1,2},
        Q. D. Wang\altaffilmark{3},
        T. Hagihara\altaffilmark{4},
        K. Mitsuda\altaffilmark{4}, 
        D. McCammon\altaffilmark{5}, and
	N. Y. Yamasaki\altaffilmark{4}
}

\altaffiltext{1}{Massachusetts Institute of Technology (MIT) Kavli Institute
 for Astrophysics and Space Research, 70 Vassar Street, Cambridge, MA 02139;
 yaoys@space.mit.edu}
\altaffiltext{2}{University of Colorado, CASA, 389 UCB, Boulder, CO 80309; yaoys@colorado.edu}
\altaffiltext{3}{Department of Astronomy, University of Massachusetts, 
  Amherst, MA 01003}
\altaffiltext{4}{Department of High Energy Astrophysics, Institute of Space and
Astronautical Science (ISAS), Japan Aerospace Exploration Agency (JAXA), 3-1-1, Yoshinodai, Sagamihara, 229-8510, Japan}
\altaffiltext{5}{Department of Physics, University of Wisconsin-Madison, 1150 University Avenue, Madison, WI 53706}

\begin{abstract}
We present {\sl Suzaku} spectra of X-ray emission in the fields just 
off the LMC~X--3 sight line. \ovii, \oviii, and \neix\ 
emission lines are clearly detected, suggesting the presence of an 
optically thin thermal plasma
with an average temperature of 2.4 $\times 10^6$ K. This temperature
is significantly higher than that inferred from existing X-ray absorption line
data obtained with \chandra grating observations of \xs, 
strongly suggesting that the gas is not isothermal. We then jointly analyze 
these data to characterize the spatial and temperature distributions of the 
gas. Assuming a vertical exponential Galactic disk model, we estimate the gas 
temperature and density at the Galactic plane and their scale heights 
as $3.6(2.9, 4.7)\times10^6$ K and $1.4(0.3, 3.4)\times10^{-3}~{\rm cm^{-3}}$ 
and  $1.4(0.2, 5.2)$ kpc and $2.8(1.0, 6.4)$ kpc, respectively. This 
characterization can account for all the \ovi\ line absorption, as observed 
in a \fuse\ spectrum of LMC~X--3, but only predicts less than one tenth of 
the \ovi\ line emission intensity typically detected at high Galactic 
latitudes. The bulk of the \ovi\ emission most likely 
arises at interfaces between cool and hot gases.

\end{abstract}


\section{Introduction }
\label{sec:intro}

The Galactic diffuse hot gas at temperatures $\sim10^6$ K can be effectively 
probed via its emission and absorption features in X-ray 
and far-ultraviolet (UV) wavelength bands.
The previous X-ray emission investigations were largely based on the 
broadband X-ray background data, e.g., {\sl ROSAT ALL Sky Survey} 
(RASS; Snowden \etal 1997). A high 
spectral resolution X-ray calorimeter abroad a sounding rocket, though 
providing little spatial resolution, clearly detected
the \ovii\ and \oviii\ emission lines, confirming that
much of the soft X-ray background (SXB) emission is thermal in origin 
\citep{mcc02}. Recently, several groups have attempted to study
the background emission with X-ray CCDs aboard {\sl Suzaku X-ray Observatory},
which, compared to those on {\sl XMM-Newton} and {\sl Chandra X-ray observatories},
have a significantly improved spectral resolution and a low instrument
background (e.g., Smith \etal 2007; Henley \& Shelton 2007). 
These later X-ray observatories, however, also carry the
high resolution grating instruments that allow for the detection of the 
X-ray absorption lines (e.g., from \ovii, \oviii, and \neix\ K$\alpha$ 
transitions) by diffuse hot gas
in and around the Galaxy. Indeed, such absorption lines 
are detected in grating spectra of nearly all Galactic and extragalactic 
sources as long as the spectral signal-to-noise ratio is high enough 
(e.g., Futamoto \etal 2004; Yao \& Wang 2005; Fang \etal 2006; 
Bregman \& Llyoid-Davis 2007). These X-ray absorption lines trace gas over 
a broad temperature range of $\sim 10^{5.5}-10^{6.5}$~K.

Gas at temperatures $\lsim 10^{5.5}$~K may be traced more sensitively in 
far-UV, via the
detection of the \ovi\ lines at $\lambda\lambda$1031.96 and 1037.62. Extensive
observations of the lines in absorption have been carried out with 
{\sl Copernicus} and {\sl Far Ultraviolet Spectroscopy Explorer} 
({\sl FUSE}; Jenkins 1978; Savage \etal 2003; Bowen \etal 2008). In addition, 
the \ovi\ lines have also been detected in emission with {\sl FUSE}, 
although the sky is only sampled at various Galactic 
latitudes (e.g., Shelton \etal 2001; Otte \& Dixon 2006). The intensity
of the $\lambda$1031.96 line, for example, ranges from 1800 to 9100 LU 
(line unit; 1 ${\rm photon~cm^{-2}~s^{-1}~sr^{-1}}$). However, 
the interpretation of the \ovi\ line(s) alone is not straight forward, because
in the collisional ionization equilibrium (CIE) state, the \ovi\ population 
sharply peaks at the intermediate temperature $\sim 10^{5.5}$~K where gas 
cools very efficiently \citep{sut93}. Thus such \ovi-bearing 
gas is expected to be rare and may preferentially reside at interfaces between 
cool gas clouds and thermally more stable hot gas. But this latter hot gas, 
which should be more abundant, distributed more widely, and effectively 
traced by the X-ray \ovii\ line,  could contribute significantly to 
the \ovi\ line as well. Currently, little is
known about the relative contributions from these two origins to the 
\ovi, either in absorption or in emission.

\begin{figure}
   \plotone{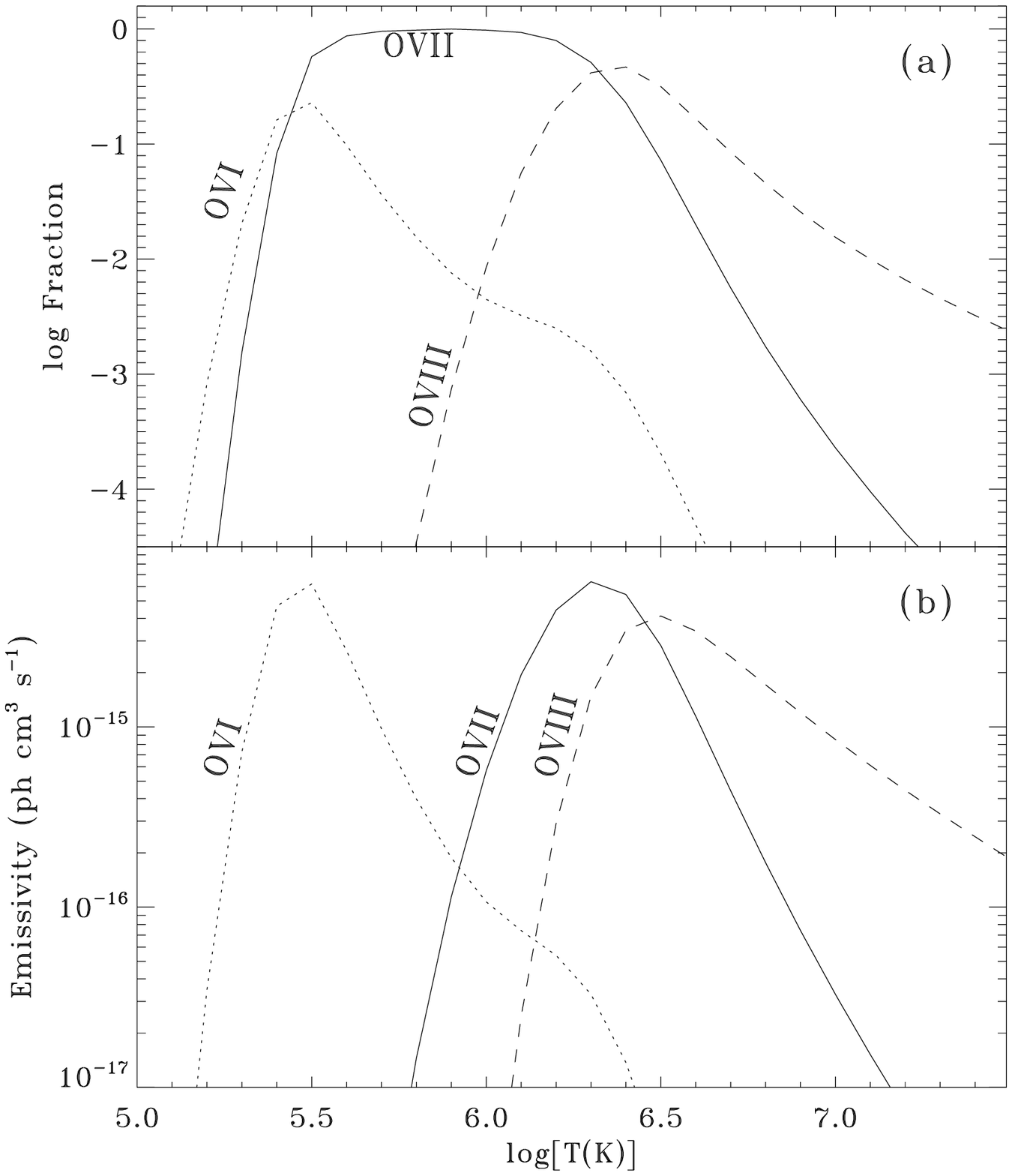}
   \caption{Ionization fraction (a) and the emissivity (b) of oxygen
	ions as a function of temperature for a gas in the collisional
	ionization equilibrium state \citep{sut93}. 
	The emissivity is a summation of the doublet transitions
	at 1031.96 \AA\ and 1037.62 \AA\ (\ovi), triplet transitions at
	22.10 \AA, 21.80 \AA, and 21.60 \AA\ (\ovii), and K$\alpha$ plus 
	K$\beta$ transitions at 18.97 \AA\ and 16.0 \AA\ (\oviii).
	The emissivity of \ovi\ is scaled down by a 
	factor of 1000 for demonstration purpose.
    \label{fig:absVSemi}
  }
\end{figure}

Clearly, a combined analysis of the X-ray and far-UV lines,
in both emission and absorption, will be the most beneficial. 
While an absorption line is proportional to the total column density of the 
gas integrated along a line of sight, an emission line depends on the emission 
measure (EM) of the gas. Furthermore, for a gas in the CIE state, 
both the ionic column density and the EM depend on the 
gas temperature, but in different manners (Fig.~\ref{fig:absVSemi}). 
Therefore a joint analysis of multiple
emission/absorption lines will enable us to constrain not only the temperature
and its distribution but also the size and density of the intervening gas 
(e.g., Shull \& Slavin 1994). When transitions from multiple elements 
(e.g., O and Ne) are detected, their relative abundances can also be 
estimated (Yao \& Wang 2006). Such a joint analysis 
has been tentatively applied to several sources 
(e.g., Yao \& Wang 2007; Shelton \etal 2007); but none of these sources has
high resolution emission and absorption data available in both X-ray and 
far-UV wavelength bands.

In this paper, we report our investigation of the hot gas along the sight line
toward LMC~X--3. \citet{wang05} have reported the detection of the
hot gas associated with our Galaxy, based on X-ray and far-UV 
absorption line spectra
from \chandra\ and \fuse\ observations. Here we present the 
{\sl Suzaku} CCD emission 
spectra of the X-ray background in two fields adjacent to
the LMC~X--3 sight line. This unique combination of the X-ray and far-UV 
spectral data toward essentially the same part of the sky further allows us to
constrain the spatial, thermal, and chemical properties of the hot gas. 

This paper is organized as follows.
We present the {\sl Suzaku} observations and the data calibration, 
as well as a brief description of the 
existing {\sl Chandra} and {\sl FUSE} data in \S~\ref{sec:obs}. 
In \S~\ref{sec:results}, we first analyze the X-ray 
emission ({\S~\ref{sec:emission})
and absorption ({\S~\ref{sec:abs}) data separately,
and then build a slab-like hot gas model to jointly analyze these 
X-ray data (\S~\ref{sec:model}). In \S~\ref{sec:fuse}, we compare the model
predicted \ovi\ absorptions with the \fuse\ detections, and then 
use the observed \ovi\
absorption line to further constrain our model. We discuss the 
implications of our results
in \S~\ref{sec:dis} and summarize our results and conclusions in \S~\ref{sec:sum}.

Throughout the paper, we assume the hot emitting/absorbing gas to be 
optically thin (see \S~\ref{sec:scattering} for further discussion) 
and in the CIE state. We adopt the solar abundances from \citet{and89}
and quote parameter errors at 90\% confidence levels
for a single varying parameter unless otherwise noted. 
We also refer the hot gas on scales of several 
kpc as the Galactic disk, in contrast to the Galactic halo on scales
 of $>10$ kpc (see \S~\ref{sec:location} for further discussion). 
Our spectral analysis uses the software package XSPEC (version 11.3.2).

\section{Observations and data reduction}
\label{sec:obs}
\subsection{{\sl Suzaku} observations}
\label{sec:suzaku}

We observed the emission of the hot diffuse gas toward off-fields of the
LMC~X--3 sight line once during the science working group (SWG) program and 
twice during the AO1 program (Table~\ref{tab:log_obs}), using the CCD camera 
XIS \citep{Koyama_etal_2007} on board {\sl Suzaku}  \citep{Mitsuda_etal_2007}.
To obtain the diffuse gas emission as close as to the LMC~X--3 sight line 
along which the X-ray \ovii\ absorption line has been detected
(see \S~\ref{sec:chandra}) but with a minimized confusion by stray lights 
from the point source, and to average out the possible spatial gradient of 
the diffuse emission intensity, we observed two fields that are in nearly 
opposite directions from LMC X--3 and $\sim30'$ away from it, 
as illustrated in Figure~\ref{fig:suzakuFOV}.
With this configuration and the roll angle of the XIS field of view, we 
estimate that stray lights from LMC~X--3 contribute no more than 6\% to the 
observed X-ray emission in 0.3--1.0 keV energy range during our observations.
The XIS was set to the normal clocking mode and the data format was either
$3 \times 3$ or $5 \times 5$, and the spaced-raw charge injection (SCI) was 
not applied to any of the data during the observations. 

\begin{figure}
   \plotone{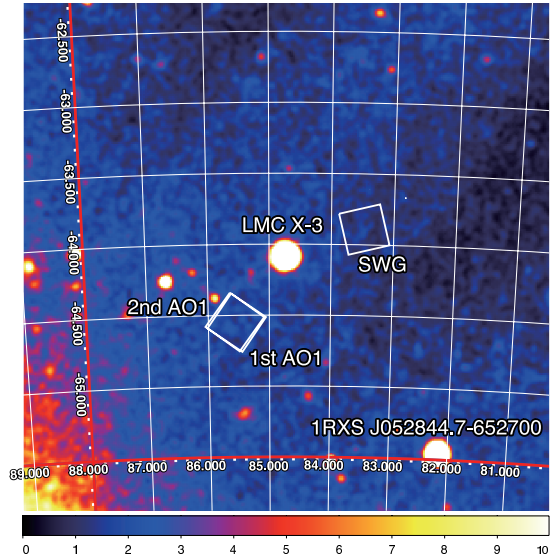}
   \caption{{\sl RASS}\ 0.1-2keV band X-ray map in the vicinity of LMC X-3 
	(the bright source at the center) and the XIS field of view of the 
	three presented observations.
    \label{fig:suzakuFOV}
  }
\end{figure}

\begin{deluxetable*}{lccc}
\tablewidth{0pt}
\tablecaption{{\sl Suzaku} observation log \label{tab:log_obs}}
\tablehead{
& SWG & 1st AO1 &2nd AO1}
\startdata
Aim point   \tablenotemark{a} (J2000)  & (83.4720,-63.9000) &(85.5500,-64.5500) &(85.5500,-64.5500) \\
Observation start  times (UT) &14:06:44, 17 March 2006&  03:27:42, 22 April
	 2006& 07:35:00, 31 October 2006\\
Observation end times (UT) &21:48:52, 19  March 2006&  00:14:19, 23 April, 2006&
		 18:01:24, 31 October 2006\\
Exposure time &  80  ks &  50  ks & compensation for 1st AO1\\
Exposure after data selection & 83.9 ks & 22.5 ks  &  14.5 ks 
\enddata
\tablecomments{$^a$ Aim point on the focal plane is the XIS nominal position.}
\end{deluxetable*}

We used processed data version 1.0.1.1 for the SWG observation and
version 2.0.6.13 for the two AO1 observations. 
We adopted the standard data selection criteria to obtain the good time
intervals (GTIs), i.e., excluding exposures when the {\sl Suzaku}'s line of 
sight is elevated above the Earth rim by less than 10$^\circ$ and
exposures with the ``cut-off rigidity'' less than 6 GV
(please refer to Smith \etal 2007 for further information).
We checked the column density of the Sun-lit atmosphere 
during the selected GTIs, and found that it is always below 
$1.0 \times 10^{15}~{\rm cm}^{-2}$, which is the criterion for no 
significant neutral oxygen emission from the Earth's atmosphere 
\citep{smith07}. We also converted the pulse-height-analysis (PHA) channels
of X-ray events to pulse-invariant (PI) ones by using the {\sl xispi} script 
(version 2006-5-16 for SWG and version 2007-05-30 for AO1 observations).

We created X-ray images in 0.3--2.0 keV energy range for 
the three observations, and found three discrete X-ray sources in the SWG 
data. To obtain the ``true'' diffuse emission, we removed those
events within circular regions centered at the sources with
a radius of $1'$ for the two faint sources and $3'$ for the
bright one. Modeling these discrete sources we estimated the 
contamination of their stray lines to the diffuse emission to be
less than 5\% in 0.3--2.0 keV energy band.



In the last step, we excluded those events severely affected 
by the solar activity. The diffuse X-ray emission below 1 keV 
could be contaminated by X-ray emission of the 
solar wind charge exchange (SWCX) if the proton flux exceeds 
$3 \times 10^8 {\rm  ~ s}^{-1} \persqrcm$ in the wind
\citep{Mitsuda_etal_2008}. The probability of such contamination 
increases if the shortest Earth-to-magnetopause (ETM) distance is
$\lsim10$ Earth radius ($R_{\rm E}$)  \citep{Fujimoto_etal_2007}.  
Here, the magnetopause is  defined as the lowest position along the line 
of sight where geomagnetic field is open to interplanetary space.  
We used solar wind data obtained with the Solar Wind Electron, Proton,
and Alpha Monitor (SWEPAM) aboard the {\it Advanced Composition Explorer} ({\it ACE}) 
\footnote{See http://www.srl.caltech.edu/ACE/ASC/level2.}.
We found that there are some time intervals with proton flux 
exceeding the above limit during the SWG and the first AO1 observations. 
We thus sub-divided the observations 
and constructed X-ray spectra for various flux levels,  
and found that, for the SWG observation, the spectra are consistent with 
each other, while for the first AO1 observation the \ovii\ and \oviii\
intensities are enhanced in the high flux subsets.
We then calculated the ETM distance, and found that during the SWG 
observation, it is always longer than 5 $R_{\rm E}$ when the magnetic field 
is open to the Sun direction and that it becomes as short as 1.4 $R_{\rm E}$ 
when the magnetic field is open to anti-Sun direction during which however 
solar-wind particles cannot penetrate. In contrast, during the first AO1 
observation the ETM distance varies significantly and some times becomes 
shorter than 3 $R_{\rm E}$ even when the magnetic field is open to the Sun 
direction. We further sorted the first AO1 data according to
this distance and the solar wind flux, 
and decided to use only those time intervals with the proton flux
$<3 \times 10^8~ {\rm s}^{-1} \persqrcm$ or the ETM distance
$>3R_{\rm E}$, during which spectra of the SWG and the AO1 data
are all consistent.

We estimated the non-X-ray background from the night Earth database 
using the method described in \citet{Tawa_etal_2008}. We found that
for XIS1, the count rate discrepancy between 
the database and our observations is about 5\% above 10 keV,  
which indicates the background uncertain level.  
Since the non-X-ray background is only 10\% of the diffuse 
emission for the energies below 1 keV, this uncertainty is negligible.
 
We constructed instrumental response files (rmf) and effective area (arf)
by running the scripts {\sl xisrmfgen} and {\sl xissimarfgen} 
(version 2006-10-26; Ishisaki \etal 2007).
To take into account of the diffuse stray light effects, we  
used a 20$'$-radius flat field as the input emission in calculating
the arf.  We also included in the arf file
the degradation of low energy efficiency due to the contamination on the XIS
optical blocking filter.

In this work, we only used the spectra obtained with XIS1, which is a
backside-illuminated CCD chip and is of high sensitivity at photon energies
below 2 keV compared to the other three 
frontside-illuminated CCDs, XIS0 and XIS2-3 
\citep{Koyama_etal_2007}. We grouped the spectra to a signal-to-noise 
ratio of $\geq10$ in each channel, and used energy range of 0.4-3.0 keV in our 
analysis. This range is broad enough for constraining the continuum 
and also covers the H- and He-like emission lines of nitrogen, oxygen, neon, 
magnesium, and L transition of \fexvii. The \nvii, \ovii, \oviii, and \neix\ 
lines are clearly visible in the spectra (Fig.~\ref{fig:suzaku}).  

\begin{figure}
   \plotone{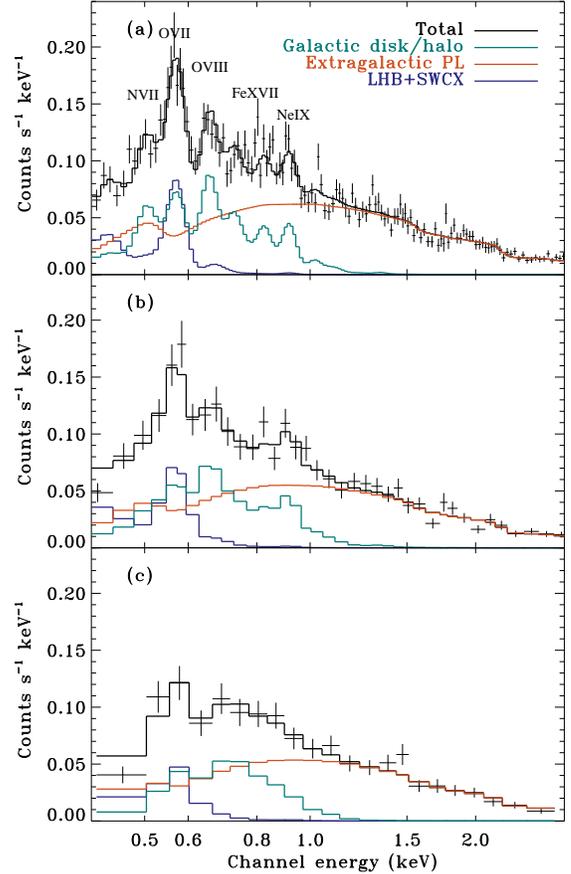}
   \caption{The background emission spectra in the off-fields of the LMC~X--3 
	sight line obtained with {\sl Suzaku} XIS1 in the SWG ({\sl a}), the 
	first AO1 ({\sl b}), and the second AO1 ({\sl c}) observations, 
	respectively. The spectra were re-grouped to be with a signal-to-noise 
	ratio of $\geq10$.
    \label{fig:suzaku}
  }
\end{figure}

The observed diffuse X-ray emission is mainly from the Galactic gas and 
confusion from the LMC is little. The targeted sight lines are well outside the
main body of the LMC and is $\sim5^\circ$ away from the active star-forming 30
Doradus region (see Fig.~1 in Wang et al. 2005). Toward these sight lines, 
there is no evidence for the present of a large-scale halo; the hot halo gas 
producing the observed \ovii\ and \oviii\ emission could be at too 
high temperatures to be confined by the LMC. Furthermore, comparing to the 
AO1 pointings, the SWG pointing is $\sim1^\circ$ degree further away from the 
LMC (Fig.~\ref{fig:suzakuFOV}). The consistent measurements between the 
observations 
(see \S~\ref{sec:emission}) indicate a lack of a radial gradient that 
could be expected in any LMC emission contribution. We therefore conclude 
that the LMC contribution to the observed emission, if any, is negligible.

\subsection{{\sl Chandra} and {\sl FUSE} observations}
\label{sec:chandra}

{\sl Chandra} and {\sl FUSE} observed LMC~X--3 with exposures of $\sim100$ 
and 120 ks, respectively. Based on these observations, \citet{wang05} reported
the detection of \ovii, \neix, and \ovi\ absorption lines along the
sight line. For ease of reference, we list these detection in 
Table~\ref{tab:line}. A joint analysis of the detected \ovii\ 
and \neix\ K$\alpha$ absorption lines, together with the non-detection lines 
of \ovii\ K$\beta$ and \oviii\ K$\alpha$, allowed for the measurements of the 
characteristic temperature, velocity dispersion, and oxygen column density of 
hot gas along the sight line with little confusion with extragalactic gas 
\citep{wang05}. We adopted the same co-added spectra 
and the corresponding calibration files as obtained in \citet{wang05}. 
Because the {\sl Chandra} observation was carried out with the high resolution 
camera (HRC) that has very little energy resolution itself, the spectrum 
therefore needs to be fitted in a broad spectral range with an 
order-combined response file in order to take into account the grating-order 
overlapping (please refer to Wang \etal 2005). To easily
implement, we further extracted the first grating order spectrum by doing
the following. We first obtained a ``global'' best fit to the spectrum over the
wavelength range of 2-30 \AA, and then subtracted the spectrum channel by 
channel by the difference of the model-predicted counts between based on the 
order-combined response file and based on the first-order response file.
We used the first-order spectrum in ranges of 12-22, and 24-29.5 \AA, 
covering the L transition of the \fexvii\ as well as the K transitions 
of the H- and He-like neon, oxygen, and nitrogen ions. To facilitate a joint 
analysis of the X-ray absorption and emission spectra with the \ovi\ absorption
line measurement, we re-wrote the {\sl FUSE} spectral 
file produced with the CALFUSE 
pipeline in a format that can be read into the XSPEC.
A Gaussian profile with a full-width-half-maximum of $20~{\rm km~s^{-1}}$ 
was used to mimic the instrumental response of the {\sl FUSE} data. 

\begin{deluxetable}{lccccc}
\tablewidth{0pt}
\tablecaption{Line measurements \label{tab:line}}
\tablehead{
 	& \ovi & \ovii & \oviii & \neix & \nvii }
\startdata	
EW (m\AA)$^a$ 	& $235.8\pm37.5$ & $20\pm6$    & $<4.1$ & $6.3\pm2.6$ & $<1.7$\\
I (LU)$^b$      & $\cdots$	 & $5.0\pm0.8$ & $2.5\pm0.4$ & $1.0\pm0.2$ & $2.9\pm2.0$   
\enddata
\tablecomments{
	$^a$ Equivalent width of the absorption lines reported in  
	\citet{wang05}, except for \nvii\ that is measured
	in this work. 
	$^b$ Line intensity (1 LU = 1 ${\rm photon~cm^{-2}~s^{-1}~sr^{-1}}$) 
	measured from the unresolved triplet of \ovii\ and \neix, and 
	K$\alpha$ of \oviii\ and \nvii\ in \S~\ref{sec:emission}.
}
\end{deluxetable}

\citet{wang05} argued that the observed absorption lines trace the 
Galactic gas rather than any intrinsic material associated with LMC~X--3. 
Here we recapitulate their argument in the following. LMC~X--3 has a 
systematic velocity of $\sim300~{\rm km~s^{-1}}$. However, with the typical 
velocity resolution of $\sim20~{\rm km~s^{-1}}$ offered by {\sl FUSE}, the 
centroid of the \ovi\ absorption line was measured as 
$55\pm11~{\rm km~s^{-1}}$ (1$\sigma$ error), which is $>10~\sigma$ smaller
than the velocity of LMC~X--3. The consistent measurements of this line in two
observations with different source flux also indicate a Galactic origin. 
In measuring the absolute wavelength, current X-ray instruments can not offer 
a significantly better resolution
than that systematic velocity, but {\sl Chandra} grating 
observations can provide a relative wavelength measurement as accurate
as several tens kilometer per second. Wang et al. found that the velocity 
shifts of the \ovii, the low ionization \oi\ K$\alpha$ lines, and the
wavelength interval between these two lines are consistent with those 
observed toward a Galactic source 4U~1820--303 and two extragalactic
sources Mrk~421 and PKS~2155--304; the lines along the latter three
sight lines are believed to have a Galactic origin. On the other hand, if 
the absorptions are associated with the LMC~X--3 as in a photo-ionized wind
scenario, the line width and velocity shift are expected to be on the order of
the escaping velocity of the system ($\sim10^3~{\rm km~s^{-1}}$), which is
inconsistent with the narrowness and rest-frame velocity of the observed
\ovii\ and \ovi\ lines. They then attributed the observed absorptions to
the Galactic gas, which is also assumed in this work.

In the {\sl FUSE} spectra of LMC~X--3, \citet{hut03} and \citet{wang05} also
found an \ovi\ emission line whose velocity centroid varies as a function of 
the binary phase. This phase dependency could be explained if the emission
arises from the companion star illuminated by the X-ray primary
(L. Song et al. in preparation). The similar X-ray \ovii\
emission has not been observed. The lines of our interests in this work are 
all very narrow (with a FWHM of $\sim10^2~{\rm km~s^{-1}}$) 
and their measurements depend on the local continuum.
The observed emission line is very broad (with a half-width 
of $\sim10^3~{\rm km~s^{-1}}$) and therefore will not affect our
measurements herein presented. 

\section{Analysis and results}
\label{sec:results}

\subsection{Fit to X-ray emission data}
\label{sec:emission}

The SXB emission is a composite of various components. Chiefly among them
are the diffuse hot gas associated with the large-scale Galactic disk/halo, 
the Local Hot Bubble (LHB), the SWCX,
and the extragalactic (primarily from unresolved AGNs) and the Galactic 
(mainly unresolved stars) discrete sources. Recent X-ray shadowing 
experiments with {\sl Suzaku} and {\sl XMM-Newton} indicate
that the combined contribution from the LHB and the SWCX can be described 
with an emitting plasma at a temperature of $\sim1.2\times10^6$ K 
with an EM of $0.0075~{\rm cm^{-6}~pc}$ \citep{smith07, gal07, hen07},
which is equivalent to the line intensities of 3.5 and 0.04 LU
for \ovii\ triplet and \oviii, respectively. The contribution from unresolved 
stellar sources is expected to be negligible at high galactic latitudes 
($|b|>30^\circ$; Kuntz \& Snowden 2001) and is not further considered in this 
work. The extragalactic emission contribution can be 
approximated as a power-law (PL) with an index of $\sim1.4$ and a 
normalization of $\sim10.9~{\rm photons~cm^{-2}~kev^{-1}~s^{-1}~sr^{-1}}$ 
at 1 keV (e.g., Hickox \& Markevitch 2006; see \S~\ref{sec:cxb} for further 
discussion). 

We decompose the SXB emission in the \xs\ off-fields by jointly fitting
the three {\sl Suzaku} emission spectra. We first fit the spectra  with
an XSPEC model combination of mekal${\rm_{LHB}}$ + 
wabs(powerlaw${\rm_{exg}}$ + vmekal${\rm_{disk})}$. 
We fix the cool gas absorption (wabs)
to be $N_{\rm H}^c=3.8\times10^{20}~{\rm cm^{-2}}$, derived from the neutral
absorption edge study \citep{page03}. We assume the Galactic diffuse hot gas
(as denoted by vmekal$_{\rm disk}$) to be isothermal, let the 
abundances of nitrogen, neon, and iron to be fitting parameters, and
fix the abundance ratios between other metal elements and oxygen to the solar 
values. We also fix the LHB component to the values mentioned above
(but see \S~\ref{sec:LHB} for further discussion) and allow the normalization
of the extragalactic PL component (powerlaw${\rm_{exg}}$) to vary in the fit 
(accounting for the cosmic variance). This simple model combination 
fits the emission data well. The fitting parameters are consistent  
among the three observations except for slightly higher 
nitrogen abundance and PL normalization values for the SWG data.
The discrepancy in the nitrogen abundance is probably due to the calibration 
uncertainty of the XIS1 for the SWG observation taken at early stage
(see \S~\ref{sec:abs}), while the high PL normalization is likely caused by 
the residual contamination (the wing of the point spread function) of the
relatively bright field source that cannot be completely excluded from the SWG 
data (\S~\ref{sec:suzaku}). We then link all fitting parameters among three 
observations except for nitrogen abundance, the PL index, and PL 
normalization, which are linked within 
the two AO1 observations but are allowed to vary between AO1 and SWG ones. 
The fit is acceptable with $\chi^2/dof=194/179$ (Fig.~\ref{fig:suzaku}).
The fitting parameters are presented in the first row of 
Table~\ref{tab:results}. By setting the abundances of nitrogen, oxygen, and 
neon in the Galactic disk component to be zero, and using Gaussian profiles 
to fit the \nvii, \ovii, \oviii\ and \neix\ lines at 0.50, 0.56, 
0.65, and 0.91 keV, we measure the intrinsic (cool-gas-absorption corrected) 
emission line intensities of the diffuse hot gas (Table~\ref{tab:line}).

\begin{deluxetable*}{lcccccccccc}
\tabletypesize{\scriptsize}
\tablewidth{0pt}
\tablecaption{
 Spectral fitting results \label{tab:results} }
\tablehead{
& T$^a$ & EM & $N_{\rm H}^b$ &  & $h_{\rm n}$ &&&& & \\
& ($10^6$ K) & 
($10^{-3}~{\rm cm^{-6}~pc}$) & 
($10^{19}~{\rm cm^{-2}}$) & 
$\gamma$ & 
(kpc) & 
N/O$^c$  & 
Ne/O$^c$ & 
Fe/O$^c$ & 
$\Gamma$$^d$ &
norm$^e$ }
\startdata
1 & 
2.4(2.2, 2.5) & 3.5(3.1, 3.9) & $\cdots$ & $\cdots$ & $\cdots$ &
$<2.7$        & 2.6(1.9, 3.4) & 1.5(1.1, 2.1) & 
1.6(1.5, 1.8) & 13(12, 14) \\
2 & 
1.3(0.8, 2.0) &
$\cdots$ & 
2.3(1.5, 3.7) & 
$\cdots$ & 
$\cdots$ & 
$<2.2$ & 
1(fix) & 
1(fix) & 
$\cdots$ & 
$\cdots$ 
\\
3 & 
3.6(2.9, 4.7) & $\cdots$      & 2.3(1.4, 3.8) & 0.5(0.1, 1.7) & 
2.8(1.0, 6.4) & $<2.1$  &
1.7(1.3, 2.3) & 0.9(0.7, 1.1) & 1.6(1.4, 1.7) & 12(11, 13) \\
4 & 
3.5(2.9, 3.9) & $\cdots$      & 2.0(1.4, 2.4) & 0.8(0.4, 1.1) & 
2.6(1.1, 5.8) & $<2.1$  & 1.7(1.3, 2.3)
& 0.9(0.7, 1.3) & 1.6(1.4, 1.7) & 11(11, 13) \\
5 & 
2.8(2.1, 3.1) & 3.0(2.6, 3.4) & $\cdots$ & $\cdots$ & $\cdots$ &
$<3.3$        & 2.0(1.2, 3.3) & 1.0(0.7, 2.1) &
1.6(1.4, 1.7) & 13(11, 14) \\
6 & 
3.0(2.6, 4.3) & $\cdots$      & 2.0(1.4, 2.8) & 0.9(0.5, 1.3) &
2.1(0.6, 6.9) & $<2.1$  & 2.2(1.7, 2.9) & 1.3(1.0, 1.6) &
1.6(1.5, 1.8) & 12(11, 13) \\
7 & 
3.5(3.0, 4.1) & $\cdots$      & 2.0(1.7, 2.3) & 0.8(0.4, 1.0) & 
2.8(1.4, 4.5) & $<1.2$  & 2.0(1.5, 2.7) 
& 0.9(0.7, 1.2) & 1.7(1.3, 2.0) & 11(10, 12) 
\enddata
\tablecomments{$^a$ For rows 1, 2, and 5, $T$ is gas temperature for the 
	isothermal case, and for rest rows, it is the gas temperature at the 
	Galactic plane $T_0$. 
  $^b$ The equivalent hydrogen column density in the hot ISM under the 
	assumption of the solar abundance of oxygen. 
  $^c$ Element abundance ratios in units of the solar values. 
  $^d$ The power-law index. 
  $^e$ The normalization of the power-law component, in unit of
	${\rm photons~s^{-1}~cm^{-2}~keV^{-1}~sr^{-1}}$. 
  	The listed $N/O$, $\Gamma$, and $norm$ are
	constrained from AO1 observations, and the values from SWG data
	are given separately in the following. 
	The horizontal dotted symbols indicate that the constraints are not 
	applicable in the fitting. See text for details.\\
  Row 1: Results from fitting the emission data 
	(\S~\ref{sec:emission}). Values of $N/O$, $\Gamma$, and $norm$
	from SWG data are
	5.9(3.7, 7.8), 1.5(1.4, 1.6), and 15.4(14.5, 16.5), respectively. \\
  Row 2: Results from fitting the absorption data (\S~\ref{sec:abs}). \\
  Row 3: Results from the joint analysis of the X-ray absorption and 
	emission data (\S~\ref{sec:model}). 
	Values of $N/O$, $\Gamma$, and $norm$ from the SWG data
	are 4.9(3.4, 6.6), 1.4(1.3, 1.5), and 15.2(14.4, 16.1) respectively. \\
  Row 4: Results from joint analysis of all observations with
	{\sl Chandra}, {\sl Suzaku}, and {\sl FUSE} (\S~\ref{sec:fuse}). \\
  Rows 5 and 6: The same as rows 1 and 4, respectively, except for allowing 
	the EM of the LHB plus the SWCX component to vary from 
	0 to 0.00113 ${\rm cm^{-6}~pc}$	(\S~\ref{sec:LHB}). \\
  Row 7: Results of using a broken-PL to approach the extragalactic emission
	(\S~\ref{sec:cxb}). 
	The $\Gamma$ value listed here is for $E<1.0$ keV, 
	and for $E\geq1.0$ keV $\Gamma=1.4(1.3, 1.5)$. 
	For SWG data, $N/O=3.4(1.5, 5.7)$ and $norm=15.2(14.7, 16.2)$. \\
}
\end{deluxetable*}


%
%

\subsection{Fit to X-ray absorption data}
\label{sec:abs}
Assuming the intervening gas to be isothermal and adopting the ISM metal 
abundances from \citet{wil00}, \citet{wang05} constrained the gas 
temperature, velocity dispersion, and 
\ovii\ (or equivalent hydrogen) column density,
by jointly analyzing the detected \ovii\ and \neix\ K$\alpha$ absorption lines
with the non-detections of \ovii\ K$\beta$ and \oviii\ K$\alpha$ lines.
Following the same procedure, we have re-fitted the spectrum reduced 
in \S~\ref{sec:chandra} and obtained nearly identical results to 
those obtained by Wang et al. (the second row in Table~\ref{tab:results}),
except for the equivalent hydrogen 
column density $N_H$ due to the different oxygen abundance
adopted here. In order to compare with the {\sl Suzaku} result
on the relative abundance of N/O, we further included 
the non-detected \nvi\ and \nvii\ K$\alpha$ absorption lines 
in our fit (Fig.~\ref{fig:N}; Table~\ref{tab:line}).
The fit gives a 95\% upper limit, N/O$<$2.2, which
is consistent with the value constrained from the {\sl Suzaku} AO1 data, but 
is lower than that inferred from the SWG data. We therefore attribute
the apparent super-solar nitrogen abundance observed in the SWG data to 
the instrumental calibration uncertainty. 

\begin{figure}
\plotone{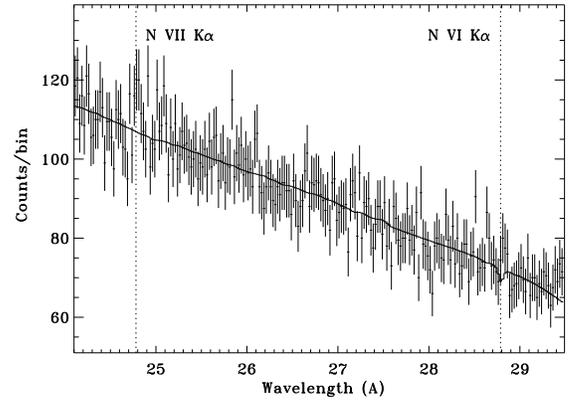}
   \caption{The {\sl Chandra}
	count spectrum around \nvi\ and \nvii\ K$\alpha$ absorption
	lines. The dotted lines mark the rest-frame wavelength positions.
	The solid line represents the best-fit continuum and the
	predicted absorptions, assuming the solar value of
	N/O. The binsize is 25 m\AA.
    \label{fig:N}
  }
\end{figure}

From now on, we let the normalization of the extragalactic PL component and 
the N/O ratio for the SWG observation vary independently from those for the AO 
ones in our data analysis. 
We will not further discuss these values in the text but still list them in 
our resulting Table~\ref{tab:results} for references.

The gas temperature inferred from the above absorption line
fits is a factor of $\sim2$ lower than that constrained from the emission 
spectral analysis (\S~\ref{sec:emission}; Table~\ref{tab:results}).  We note 
that the temperature would be even lower if a super-solar Ne/O value as 
indicated in the emission spectral analysis (\S~\ref{sec:emission})
were adopted in our absorption line fits. This temperature 
inconsistency clearly indicates that the X-ray emitting/absorbing gas
is not isothermal. In this case, the emission and absorption arise 
preferentially in different temperature ranges. 

\subsection{Non-isothermal model and joint fit}
\label{sec:model}

Motivated by the observed morphology of diffuse X-ray emission around 
the nearby disk galaxies, \citet{yao07} constructed a non-isothermal 
model for the Galactic disk hot gas. In the following, we first 
briefly formulize this model and then constrain it using the obtained
absorption and emission data along the \xs\ sight line.

Assuming that the hydrogen number density and temperature of the hot gas
can be characterized as 
\begin{equation} \label{equ:expHT}
  n  =  n_0 e^{-z/(h_{\rm n}\xi)}~{\rm and}~
  T  =  T_{\rm 0} e^{-z/(h_{\rm T}\xi)},
\end{equation}
where  $z$ is the vertical distance away from the Galactic plane, 
$n_{\rm 0}$ and $T_{\rm 0}$ are the mid-plane values, 
and $h_{\rm n}$ and $h_{\rm T}$ are the scale heights,
and $\xi$ is the volume filling factor that is assumed to be 1 in the paper,
we can derive
\begin{equation} 
n=n_{\rm 0}(T/T_{\rm 0})^\gamma,
\end{equation}
where $\gamma=h_{\rm T}/h_{\rm n}$.
Therefore, the differential 
hydrogen column density distribution is also a power law function of $T$,
\begin{equation} \label{equ:PL}
    dN_H =  n dL 
       =  \frac{N_{\rm H}\gamma}{T_{\rm 0}} (T/T_{\rm 0})^{\gamma-1} dT,
\end{equation}
and the corresponding ionic column density along a sight line with a
Galactic latitude $b$ can  be expressed as
\begin{equation} \label{equ:NH}
  N_{\rm i} = \frac{N_{\rm H} \gamma A_{\rm e} }{T_{\rm 0}}
  \int^{T_{\rm 0}}_{T_{\rm min}} 
  \left( \frac{T}{T_{\rm 0}} \right)^{\gamma-1} f_{\rm i}(T)dT,
\end{equation}
where $N_{\rm H}=n_{\rm 0} h_{\rm n}\xi/\sin b$, and 
$A_{\rm e}$ and $f_{\rm i}(T)$ are the element abundance 
and ionization fraction for the ion, and $T_{\rm min} =10^5$ K
is the minimum temperature assumed (line emission and absorption 
below this temperature are negligible; Fig.~\ref{fig:absVSemi}).

Similarly, the emission intensity (from a single line or a continuum) is 
\begin{equation} 
\label{equ:eline}
  I  = 
	\frac{A_{\rm e}}{4\pi} \int_0^L \Lambda(T) n_e n dL 
  = \frac{A_e}{4\pi}\int^{T_{\rm 0}}_{T_{\rm min}}
	\Lambda(T) \frac{dEM}{dT}dT,
\end{equation}
where $\Lambda(T)$ is the corresponding emissivity of the hot gas. The
differential EM is  
\begin{equation} \label{equ:EMpl}
  \frac{dEM}{dT} = \frac{1.2 N_{\rm H}^2 \gamma}{T_{\rm 0 }L}
  \left( \frac{T}{T_{\rm 0}} \right)^{2\gamma-1},
\end{equation}
where the factor 1.2 accounts for the helium 
contribution to the electron density, and $L=h_{\rm n}\xi/\sin b$, is the
effective pathlength of the hot gas.

Both the emission and absorption data can be used independently to constrain 
the parameters $N_H$, $T_0$, $\gamma$, and the relative element abundances 
(e.g., Ne/O; Eqs.~\ref{equ:NH} and \ref{equ:eline}). A simultaneous fit to the
emission and absorption further allows for the estimate of the $L$ parameter 
(Eq.~\ref{equ:EMpl}). We have revised our 
absorption line model developed in \citet{yao05} and the thermal emission 
model $cevmkl$ in XSPEC to facilitate the use of the ionic column density 
$N_i$ and the EM as a power law function of $T$ 
(Eqs.~\ref{equ:NH}, \ref{equ:eline}, and \ref{equ:EMpl}; Yao \& Wang 2007). 

These revised models are then used to jointly fit the absorption and emission 
data. For the absorption data, the fit includes not only the significantly
detected \ovii\ and \neix\ K$\alpha$ lines, but also the key non-detections
of the \ovii\ K$\beta$, \fexvii\ L, and K$\alpha$ transitions of
\oviii, \nex, \nvii, and \nvi\ at 18.626, 15.010, 18.967, 12.134, 24.779, 
and 28.787 \AA, respectively, which provide useful constraints
on the physical and chemical properties of the absorbing gas.
The revised  $cevmkl$ model is used to fit the Galactic disk component
of the emission spectra (vmekal$_{\rm disk}$ in \S~\ref{sec:emission}). 
The abundance ratios of N/O, Ne/O, and Fe/O, which are linked together
between the absorption and emission models, are allowed to vary in our fit.
This joint fit is satisfactory; the directly constrained model parameters 
are reported in the third row of Table~\ref{tab:results}. We then infer
the temperature scale height as $h_T=1.4(0.2, 5.2)$ kpc and the gas density 
at the
Galactic plane as $n_0=1.4(0.3, 3.4)\times10^{-3}~{\rm cm^{-3}}$.  
The confidence contours of $h_n$, $T_0$, and $N_H$ vs. 
$\gamma$ are plotted in Figure~\ref{fig:contours}.

\begin{figure}
   \plotone{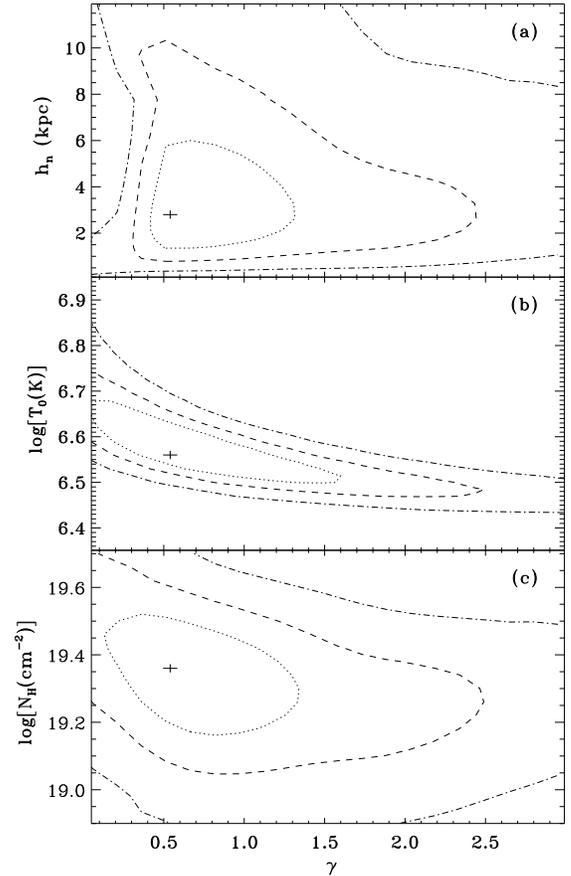}
   \caption{The 68\%, 90\%, and 99\% confidence contours of $h_{\rm n}$,
	$T_{\rm 0}$, and $N_{\rm H}$ vs. $\gamma$, obtained in a joint
	fit to the X-ray absorption and emission data (\S~\ref{sec:model}).
    \label{fig:contours}
  }
\end{figure}

\subsection{Comparison with the \ovi\ $\lambda$1031.96 absorption line}
\label{sec:fuse}

From the non-isothermal Galactic disk model constrained with the X-ray 
emission and absorption
data, we estimate the total \ovi\ contained in the diffuse hot gas as
$N_{\rm OVI} \sim 3.4\times10^{14}~{\rm cm^{-2}}$, which is consistent with
that inferred from the observed \ovi\ $\lambda$1031.96 line absorption
(Fig.~\ref{fig:fuse}; see also Wang \etal 2005; \S~\ref{sec:chandra}).
Including the \ovi\ line in our joint analysis tightens
the constraints on the model parameters 
(the forth row of Table~\ref{tab:results}), especially on the 
temperature distribution.

\begin{figure}
\plotone{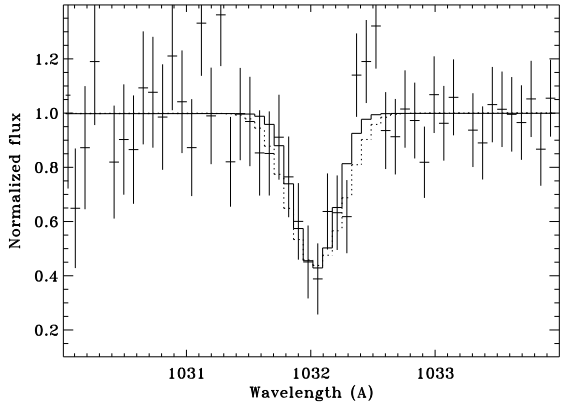}
   \caption{\ovi\ absorption line at 1031.96 \AA\ observed with {\sl FUSE}
	and the best-fit model ({\sl solid histogram}). The dotted histogram 
	represents the predicted absorption from the X-ray data constrained
	non-isothermal disk model.
    \label{fig:fuse}
  }
\end{figure}

\section{Discussion}
\label{sec:dis}

We have presented the high quality {\sl Suzaku} emission 
observations of the diffuse hot gas toward two off-fields of the 
LMC~X--3 sight line. In particular, the \ovii\ and 
\oviii\ emission lines are clearly resolved. Modeling these emission
spectra yields a
gas temperature that is about two times higher than that inferred from the 
high resolution absorption data. We find that our non-isothermal 
thick Galactic gaseous disk model can account for this discrepancy as
well as the far-UV \ovi\ absorption data. A joint fit to these data indicates 
that both the X-ray absorption and emission and the far-UV absorption 
are consistent with being
produced from hot gas in a region of several kpc around the Galactic plane.
In the following, we first discuss how our results are potentially
affected by caveats in our data analysis, and then discuss the 
implications of our results on the origin and cooling 
of the \ovii- and \ovi-bearing gases.

\subsection{Uncertainty of the LHB and the SWCX}
\label{sec:LHB}

The biggest uncertainty in modeling the SXB emission is 
the estimate of the contributions from the LHB and SWCX. 
Our understanding for both components is still very poor because it is very
hard to decompose their contributions. The SWCX emission could in 
principle be estimated by tracing the solar wind flux and the neutral atoms 
in the Earth's atmosphere and in the local ISM (e.g., Lallement \etal 2004). 
However, such an estimation is model dependent. 
For instance, for the line of sight with Galactic coordinates 
$(l, b)=(278.65^\circ, -45.30^\circ)$ observed on 2003 March 3, 
the \ovii\ emission is estimated to be 0.83 and 3.7 LU by 
\citet{kou07} and by \citet{hen08}, respectively. Recent comparisons of the
observed X-ray emission with models for the SWCX suggest that the foreground 
emission in the shadowing experiments can be mainly attributed to 
the SWCX and that the LHB emission is negligible (Koutroumpa \etal 2007; 
Rocks \etal 2007; but also see Shelton 2008). The 
total \ovii\ intensity of the SXB toward a high Galactic latitude sight line 
($l, b = 272^\circ, -58^\circ$) is 2.7(2.0, 3.4) LU (1$\sigma$ range), 
obtained from a recent {\sl Suzaku} observation \citep{mit06}.
If emission from both the LHB and the SWCX is isotropic and independent of 
Galactic latitude, this value can be regarded as the upper limit of the 
combined contribution of these two components.

Our results are not qualitatively affected by above 
discussed uncertainties. To be more quantitative in assessing the effect,
we allow EM of the LHB plus SWCX component to vary from 
0 to 0.0113 ${\rm cm^{-6}~pc}$ (\S~\ref{sec:emission}), representing two
extreme cases. The low boundary corresponds {\it zero} emission of the
LHB and the SWCX components, whereas the high boundary corresponds to the
3$\sigma$ upper limit of the observed \ovii\ intensity (i.e., 4.8 LU) 
toward the high Galactic latitude sight line. We then 
re-perform our spectral analysis described in 
\S\S~\ref{sec:emission} and \ref{sec:fuse}. We find that 
new constrained gas temperature in modeling the emission data alone under the 
isothermal assumption is still about two times higher than that constrained 
from the absorption data (refer to the first and the fifth rows 
in Table~\ref{tab:results}), which still 
necessitates the non-isothermal model developed in \S~\ref{sec:model}.
The new results constrained in the joint analysis are also consistent
with the old values, except for with a slightly broader uncertain range
(refer to the forth and the sixth rows in Table~\ref{tab:results}).



\subsection{Uncertainty of the extragalactic power-law component}
\label{sec:cxb}

In modeling the emission spectra, we used a simple PL function to approximate 
the extragalactic contribution (\S~\ref{sec:emission}). Recent {\sl Chandra} 
and {\sl XMM-Newton} deep surveys for the cosmic X-ray background
have resolved $\gsim80\%$ of the background emission at 1-8 keV into
extragalactic discrete sources (e.g., Hickox \& Makevitch 2006). 
Optically bright sources tend to be spectrally hard while faint sources tend 
to be spectrally soft (e.g., Mushotzky \etal 2000). 
To examine the goodness of our 
approximation, we replace the simple PL with a broken PL and fix the break 
energy at 1 keV to model the extragalactic component in our joint 
analysis described in \S~\ref{sec:model}. We obtain the photon indices
as 1.7(1.3, 2.0) and 1.4(1.3, 1.5) below and above 1 keV, respectively,
and find that the parameters of the Galactic hot gas are barely affected
(the seventh row of Table~\ref{tab:results}). 

\subsection{Effect of resonant scattering of the \ovii\ line emission} 
\label{sec:scattering}

In this work, we have assumed the emitting/absorbing gas to be optically thin,
i.e., we have neglected the resonant scattering effect of the hot gas. To 
qualitatively assess the validity of our assumption, we assume the hot gas 
density to be uniform for easy of analysis. The observed \ovii\ emission line 
consists of unresolved triplet transitions, resonance, forbidden, and
intercombination lines at 21.60 \AA, 22.10 \AA, and 21.80 \AA, respectively.
Since the oscillation strength is essentially zero for the latter two
transitions, we consider the ``scattering'' only for the resonant line.
The absorbed resonant photons should be re-emitted isotropically;
half of these re-emitted photons are expected to favor the observing 
direction. Taking all these into account, 
we obtain the observed line intensity as 
\begin{equation}
I = (1 - f) I_0 + 0.5fI_0 \left( 1 + \frac { 1 - e^{-\tau} }{\tau} \right ),
\end{equation}
where $\tau$ is the (max) absorption optical depth at center wavelength
of the resonant transition, and
$I_0$ is the ``intrinsic'' (without the scattering) line intensity. Here, 
$f$ is the resonant transition fraction of the \ovii\ triple, which is a 
function of gas temperature. Taking the (max) temperature at the Galactic
plane $T_0\sim3\times10^6$ K (Table~\ref{tab:results}), which favors 
more to the resonant transition and gives $f=0.65$, we get 
$I \gsim 0.675I_0$ for any value of $\tau$. This indicates that the observed
intensity should not be different from the intrinsic value by a factor 
larger than 1.5. In reality, the ``background'' gas cloud also scatters the 
photons emitted by the ``foreground'' cloud to the observing direction. If 
this effect is further considered, the difference should largely disappear.
Therefore we conclude that the resonance scattering should not significantly 
affect our results, although a more detailed modeling needs to be performed
to account for a more realistic geometry of the hot gas distribution.

\subsection{Origin of the \ovii-bearing gas in general}
\label{sec:location}

Our analysis indicates that the gas responsible for the observed  
X-ray absorption/emission along the LMC~X--3 sight line
has a pathlength about a few kpc and is consistent with 
the Galactic disk in origin. This 
result has important implications for understanding the structure and
origin of hot gas around our Galaxy.  As described in \S~\ref{sec:intro}, 
high ionization X-ray absorption lines with zero velocity shift
have been observed along 
many extragalactic sight lines, and the \ovii\ emission lines 
are also detected at various high 
Galactic latitudes. Several scenarios have been proposed for the origin 
of the emitting/absorbing gas, including the intergalactic medium 
(IGM) of the Local Group, the large-scale Galactic halo, and the thick
Galactic gaseous disk (see Yao \& Wang [2007] for a review).
These scenarios cannot be distinguished kinematically because of the limited 
spectral resolution of current X-ray instruments. So currently the 
most direct information about the location of the gas comes from 
differential analysis of the X-ray absorption lines toward sources at
different distances. \citet{wang05} find that the \ovii\ absorption 
along the sight line toward LMC~X--3, at a distance of $\sim50$ kpc, is 
comparable to those observed toward AGN sight lines.
Therefore, if the LMC~X--3 sight line is representative, one can then
conclude that the bulk of the X-ray absorption around the Galaxy is
within $\sim50$ kpc. A similar conclusion has been reached based on 
the estimate of the total baryonic matter contained in 
the \ovii-bearing gas, and on the angular distribution of the 
\ovii\ absorption \citep{fang06, bre07}. 
A comparison of a Galactic sight line (4U~1957+11)
with two extragalactic sight lines toward LMC~X--3 and Mrk~421 
further indicates that there is
no significant \ovii\ absorption in the extended Galactic halo beyond
a few kpc \citep{yao08}, which is consistent with the conclusion drawn in
this work.

The \ovii\ emission from the hot gas beyond LMC~X--3 is not expected to be
important either. 
\citet{yao08} obtained a 95\% upper limit to the \ovii\ absorption
beyond LMC~X--3 as $N{\rm _{OVII}<3.7\times10^{15}~cm^{-2}}$. Assuming this
much of \ovii\ uniformly distributed in a region 
with an extension scale of $l$, we calculate its emission as 
$I_{\rm OVII}<0.025/(A_{\rm O0.1}l_{\rm 100kpc})$ LU for the
entire temperature range from $10^5$ to $10^7$ K, where 
$A_{\rm O0.1}$ is the oxygen abundance in unit of 10\% solar value and
$l_{\rm 100kpc}$ is in unit of 100 kpc. In contrast, the inferred the \ovii\
intensity of the Galactic disk diffuse gas is 5.0 LU 
(\S~\ref{sec:emission}; Table~\ref{tab:line}).

We have shown that the X-ray emission and absorption as well as 
the far-UV \ovi\ absorption along the
sight line toward \xs\ can be explained by the presence of a thick 
Galactic hot gaseous disk. A similar explanation holds
for the sight line toward Mrk~421, for which another 
joint X-ray absorption/emission and UV-absorption has been
carried out (the X-ray emission is, however, mostly based on the RASS
broad-band measurements; Yao \& Wang 2007). The characteristic
scale height of the disk is also consistent with the \neix\ column 
density distribution in the Galaxy \citep{yao05}. The gas in the disk is
shown to have a normal metal abundances \citep{wang05, yao06}
and is clearly due to the heating by mechanical energy feedback from 
stars in the Galaxy (e.g., Ferri\`ere [1998]). Such an interpretation is 
also supported by the X-ray emission observations of 
nearby disk galaxies like our own. The diffuse X-ray emission has been
routinely observed in many late-type normal star-forming spirals and
is confined to a region around galaxies with vertical extent no more than
a few kpc (e.g., Wang \etal 2001, 2003; T\"ullmann \etal 2006a, 2006b).
Both the intensity and the spatial extent appear to be scaled with
galactic star formation rate (e.g., T\"ullmann \etal 2006b). Therefore,
we conclude that the X-ray emitting/absorbing and \ovi-absorbing gas,
as studied in this paper, arises primarily from the stellar feedback
in the Galactic disk.

\subsection{Location of the \ovi-bearing gas}
\label{sec:o6location}

Before discussing the \ovi-bearing gas location, let us look at the \ovi\ 
emission properties first. In \S~\ref{sec:fuse} we have shown that the 
observed \ovi\ absorption can be predicted from the X-ray-data-constrained
non-isothermal gas model and have then 
used the \ovi\ $\lambda$1031.96 absorption line to further constrain  
our spectral model. With the fitted model parameters, we can estimate
the intrinsic \ovi\ emission intensity of the diffuse hot gas toward LMC~X--3, 
and then compare it with observations. The best-fit model gives
186 LU in total for the \ovi\ doublet $\lambda\lambda$1032 and 1038.
Taking all the 90\% boundaries that favor more \ovi\ emission,
we obtain a firm upper limit of 1612 LU.

There is no direct measurement of the \ovi\ emission within 1$^\circ$ of the 
LMC~X--3 sight line. Within $\sim5^\circ$, some observations show measurable 
\ovi~$\lambda1032$ emission as $2.1-5.7 \times10^3$ LU, while other  
observations only yield upper limit even with longer exposures 
\citep{ott06, dix06, dix08}. Since the \ovi\ emission could vary by a factor 
of $>2.3$ at an angular 
scale as small as $25'$ \citep{dix06}, it is nearly impossible to reliably
estimate the \ovi\ emission toward LMC~X--3 sight line based on the few 
existing nearby samples. However, if there were a measurable \ovi\ emission 
toward
the LMC~X--3 sight line (but see below), it must be in order of several 
thousand LU, as typically observed at high galactic latitudes, which is  
significantly larger than that predicted in the \ovii-bearing gas
and therefore unlikely arises from the diffuse hot gas.

Where is the \ovi-bearing gas then? To answer this question, it 
is important to
compare the spatial distributions of the \ovi\ absorption and emission.
In this work, we find that most of the \ovi\ absorption can arise from the 
diffuse gas along a vertical scale of $\sim$3 kpc around the Galactic plane.
Figure~\ref{fig:OVI_dis} shows the density and column density
distributions of \ovi\ and \ovii\ as a function of the vertical distance,
predicted from our constrained non-isothermal disk model.
While compared to \ovii, \ovi\ is distributed in a relative 
narrow region, which is mainly due to the sensitive dependence
of \ovi\ population on temperature (Fig.~\ref{fig:absVSemi}),
about 80\% of the \ovi\ column still spreads over as large as 1 
(from 2.7 to 3.7) kpc. Observationally, \ovi\ absorption has been 
detected toward essentially all sight lines, and 
the characteristic scale height of the \ovi\ column densities
is similar to that constrained in this work (e.g., Savage \etal 2003). 
It has been
suggested that \ovi\ could largely arise at interfaces between hot and
cool interstellar media. However, for Mrk~421 sight line,
\citet{sav05} counted the number of cool gas velocity components 
and found that the interfaces can only account for at most 50\% of \ovi\ 
absorption, leaving a bulk of the \ovi\ to other origins. In contrast, 
the \ovi\ emission is measured toward only $\sim30\%$
of the surveyed sky \citep{ott06, dix08}. 
These results clearly indicate that the ``commonly'' observed \ovi-absorbing 
gas in kpc scale is not primarily responsible for the detected \ovi\ emission.
We propose that the low temperature regime of the kpc-scale diffuse gas could 
contain a significant or even dominant fraction of the observed \ovi\ 
absorption, as indicated in this work, and that the conductive interfaces
between hot and cool gases, though could be another 
reservoir of \ovi\ ion, is mainly responsible 
for the observed \ovi\ emission. This picture not only explains the spatial 
distribution of the \ovi\ absorption, but also naturally explains the 
sight line to sight line \ovi\ variation in both absorption and emission.
For LMC~X--3 sight line in particular, since all the observed \ovi\ 
can be attributed to the diffuse gas, the \ovi\ emission is not
expected to be strong.

\begin{figure}
\plotone{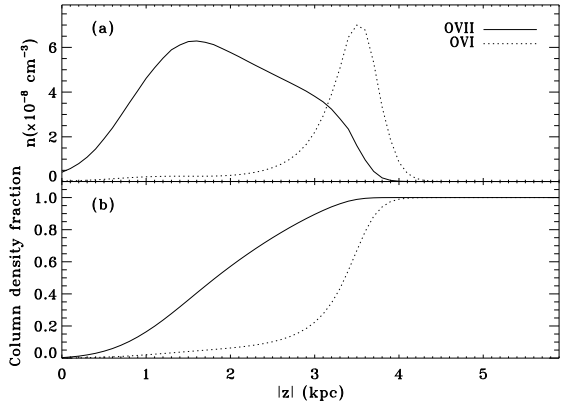}
   \caption{Spatial distribution of density (a) and column density (b) of 
	\ovi\ and \ovii, predicted
	with our constrained non-isothermal disk model. 
    \label{fig:OVI_dis}
  }
\end{figure}

\subsection{Radiative energy loss rate of the hot gas}
\label{sec:lossrate}

If hot gas toward the LMC~X--3 sight line represents reasonable well the hot 
ISM of the Galactic disk as a whole, we can then use our non-isothermal gas 
model to estimate the total radiative energy loss rate of the gas, and compare
it to the expected energy input from SN explosions. It has long been observed
that in the low $L{\rm_x}/L{\rm_B}$ ($L_{\rm B}$ is the blue luminosity)
early-type galaxies, X-ray radiative energy loss rate of the diffuse hot 
gas is far less than the expected SN energy input
(e.g., Canizares \etal 1987; Brown \& Bregman 2000). However, all the
previous studies were based on imaging observations in which decomposing
different emission components is a complicated task (e.g., Li \etal 2007)
and could make the inferred properties of the ``true'' diffuse hot gas very 
uncertain. The LMC~X--3 sight line is so far the only sight line along which
the high resolution X-ray emission/absorption and far-UV absorption 
measurements with a minimal confusion are available, which allow us to tightly
constrain the thermal and chemical properties as well as the global spatial 
distribution of the 
Galactic diffuse hot gas, as presented in this work. It is thus worthwhile
reexamining the radiative power of the Galactic hot gas based on these 
characterizations.

The energy loss rate of the hot gas can be expressed as
\begin{equation} \label{equ:lossrate}
    \frac{d^2U}{dsdt} = 2\times\int_0^\infty  1.2n^2\Lambda(T)~dz,
\end{equation}
where the factor 2 accounts for the two sides of the Galactic disk.
With our model (Eq.~\ref{equ:expHT}), we can rewrite the 
Eq.~\ref{equ:lossrate} as 
\begin{equation} \label{equ:rate}
 \frac{d^2U}{dtds} = 2\times\int^{T_0}_{T_{min}} 1.2n_0^2
	\left(\frac{T}{T_0}\right)^{2\gamma-1}\Lambda(T)
	(-h_T)~d\left(\frac{T}{T_0}\right). 
\end{equation}
Assuming the solar abundances, plugging in the best-fit $n_0$, $T_0$, 
$\gamma$, and $h_T$ values (\S~\ref{sec:model}), and taking the 
emissivity $\Lambda(T)$ from the 
atomic database ATOMDB\footnote{http://cxc.harvard.edu/atomdb},
we obtain the local surface emissivity of the hot gas as 
3.1 and 16.1 $\times10^{36}~{\rm ergs~s^{-1}~kpc^{-2}}$ in 
0.1-10 and 0.01-10 keV bands, respectively. Including the additional 
\ovi\ emission of $\sim$3000 LU from the interface component, 
the total radiative energy loss rate is
$\sim3\times10^{37}~{\rm ergs~s^{-1}~kpc^{-2}}$.

The radiative cooling of the Galactic disk hot gas only accounts for
less than 10 per cent of the expected energy input of the stellar feedback.
At Sun's galactocentric radius, the SN rate is 19 and 2.6 
${\rm Myr^{-1}~kpc^{-2}}$ for type II and type Ia, respectively
\citep{fer98}. If on average each type II SN progenitor releases 
$2\times10^{50}$ ergs of energy before it explodes and each SN explosion 
releases $10^{51}$ ergs \citep{fer98, lei92}, the total energy input is then
$8\times10^{38}~{\rm ergs~s^{-1}~kpc^{-2}}$, which is about
20 times higher than our estimated gas cooling rate.
The ``missing'' energy could be either emitted in other wavelength bands 
(e.g., infrared) or have been consumed in driving other Galactic activities
(e.g., galaxy-size out flows).

Obviously, such an estimation is model dependent.
\citet{she07} recently concluded that $\sim70\%$ of the SN energy input could
be radiated away by the Galactic hot gas. In their model, they required the
observed \ovi\ emission to be co-spatial with the \ovi\ absorption and assumed
both the \ovi-bearing and the hotter \ovii-bearing gas to be isobaric. This
requirement results in that most of the \ovi-bearing gas is confined 
within a region
of $<100$ pc in size, which favors to produce about ten times more thermal 
emission in the energy range of 0.01-0.1 keV than our model. 
In contrast, we 
constrain our model with X-ray \ovii\ and \oviii\ emission and absorption,
and find that the most of the \ovi\ absorption could arise from the
kpc-scale diffuse gas. In our model, because of large 
extent of the diffuse gas (therefore a low density), its emitting 
power is at most as much as that of the observed/expected \ovi\ 
emission lines, which presumably arise from the interfaces between
the hot and cool interstellar media (\S~\ref{sec:o6location}).
The scale size of the \ovii- and \ovi-bearing gas 
derived in this work and utilized in our estimation,
is consistent with that obtained in the recent \ovi\ absorption surveys 
\citep{sav03, bow08}.

\subsection{Evidence for oxygen and iron depletion?}
\label{sec:depletion}

In our analysis, we find that, while the relative abundance of Fe/O is 
about the solar value, the Ne/O is about 2 times higher
(Table~\ref{tab:results}). This overabundance of neon, if true, 
is consistent with the interpretation of about 50 per cent of oxygen and
iron being depleted into dust grains. However, because this overabundance
is mainly derived from the emission data in which decomposing various
components is still very uncertain, the depletion interpretation
therefore may not be unique. For instance, \citet{hen08} suggest that the
SWCX could also produce apparent non-solar abundance ratio. Furthermore, our 
understanding of the chemical abundances in the solar system is still poor.
For example, the oxygen value has recently been revised downward by $\sim35\%$ 
\citep{asp05}, but this revision is still under debate \citep{ant06}.
The abundance pattern in the ISM 
is also suggested to be slightly different from that in the solar system
\citep{wil00}. So it is more useful to list the 
number density ratio of Ne/O derived in this work, which is 0.25(0.19, 0.33).
We prefer to defer the interpretation of the apparent overabundance of neon 
until a better understanding of both the SWCX and the solar chemical 
abundances is reached.

\section{Summary}
\label{sec:sum}

We have presented an extensive study of diffuse hot gas along the
sight line toward LMC~X--3, based on new {\sl Suzaka}  X-ray CCD observations
of the background emission as well as high resolution
X-ray and far-UV absorption data. Our main results and
conclusions are summarized in the followings:

1. We have detected \ovii, \oviii, and \neix\ emission lines in the {\sl Suzaku}
emission spectra. Modeling these emission data gives a characteristic 
temperature for the emitting gas as $2.4(2.2, 2.5)\times10^6$ K, assuming
the gas to be isothermal and in the collisional ionization equilibrium.
This temperature is about two times higher than that inferred from
the X-ray absorption data under the same assumptions.
Adopting different neon to oxygen abundance ratio in
analyzing the absorption data will not relieve this discrepancy.
This inconsistency indicates the non-isothermality of the 
emitting/absorbing gas.

2. We find the X-ray emission/absorption data are consistent with
a non-isothermal hot gas model, in which both the 
gas temperature and the gas density decrease exponentially with the 
vertical distance away from the Galactic plane. Jointly analyzing the
X-ray emission and absorption data, we have constrained
the scale heights and the middle plane values of the
gas density and temperature as 
2.8(1.0, 6.4) kpc and $1.4(0.3, 3.4)\times10^{-3}~{\rm cm^{-3}}$, 
1.4(0.2, 5.2) kpc and $3.6(2.9, 4.7)\times10^6$ K, respectively. 
These scale heights indicate that the bulk of \ovii\ absorption observed in the
extragalactic sources (e.g., Mrk~421) is consistent with a Galactic in origin,
although additional relatively small 
extragalactic contribution can not be ruled out.
 
3. We find that the above X-ray constrained non-isothermal hot gas model
can naturally explain all the observed \ovi\ line absorption, but only
accounts for less than 10 per cent of the \ovi\ emission typically observed
at high galactic latitudes. We propose that most of the \ovi\ absorption 
arises from the kpc-scale diffuse gas
and that the \ovi\ emission arises chiefly from interfaces between
hot and cool media.

4. The cooling of the Galactic disk hot gas described here is only 
responsible for less than 10 per cent of the expected 
total SN energy input. The bulk of the SN energy 
is thus ``missing'' and may be radiated away in other wavelength 
bands or may have been driven into a large-scale low-density
Galactic halo or into the intergalactic space.



\acknowledgements 
We gratefully acknowledge extensive conversations with Blair Savage 
on the resonant scattering issue and conversations with Robin Shelton 
on the emission from the Local Hot bubble and the SWCX. 
We are also grateful to Claude Canizares for useful discussions
and the anonymous referee for comments, which helped to clarify our
presentation. This work is supported by NASA through the 
Smithsonian Astrophysical Observatory contract SV3-73016 to 
MIT for support of the Chandra X-Ray Center under contract NAS 08-03060
and NASA/GSFC grant NNX07AB06G to University of Massachusetts.

\clearpage

\clearpage

\clearpage

\clearpage


\begin{thebibliography}{}

\bibitem[Anders \& Grevesse (1989)] {and89} Anders, E., \& Grevesse, N. 1989, Geochim. Cosmochim. Acta, 53, 197 
\bibitem[Antia \& Basu (2006)]{ant06} Antia, H. M., \& Basu, S. 2006, \apj, 644, 1292
\bibitem[Asplund \etal (2005)]{asp05} Asplund, M., Grevesse, N., \& Sauval, J., 2005, ASPC, 336, 25
\bibitem[Birnboim \& Dekel (2003)]{bir03} Birnboim, Y, \& Dekel, A. 2003, \mnras, 345, 349
\bibitem[Brown \& Bregman (2000)]{bro00} Brown, B., \& Bregman, J. N. 2000, \apj, 539, 592
\bibitem[Bowen \etal (2008)]{bow08} Bowen, D. V., \etal 2008, \apjs, 176, 59
\bibitem[Bregman \& Lloyd-Davies (2007)]{bre07} Bregman, J. N., \& Lloyd-Davies, E. J. 2007, \apj, 669, 990 
\bibitem[Burrows \& Mendenhall (1991)]{bur91} Burrows, D. N., \& Mendenhall, J. A. 1991, Nature, 351, 629
\bibitem[Canizares \etal (1987)]{can87} Canizares, C. R., Fabbiano, G., \& Trinchieri, G. 1987, \apj, 312, 503
\bibitem[Dixon \etal (2006)]{dix06} Dixon, W. V. D., Sankrit, R., \& Otte, B. 2006, \apj, 647, 328
\bibitem[Dixon \& Sankrit (2008)]{dix08} Dixon, W. V. D., \& Sankrit, R. 2008, \apj, in press, astro-ph/0807.1237
\bibitem[Fang \etal (2006)]{fang06} Fang, T., \etal 2006, \apj, 644, 174 
\bibitem[Ferri\`ere (1998)]{fer98} Ferri\`ere, K. 1998, \apj, 497, 759
\bibitem[Fujimoto et al. (2007)]{Fujimoto_etal_2007} Fujimoto,R.  et al. 2007, \pasj, 59, S133
\bibitem[Futamoto \etal (2004)]{fut04} Futamoto, K., \etal 2004, 605, 793 
\bibitem[Galeazzi \etal (2007)]{gal07} Galeazzi, M., \etal 2007, \apj, 658, 1081
\bibitem[Henley \etal (2007)]{hen07} Henley, D. B., Shelton, R. L., \& Kuntz, K. D. 2007, \apj, 661, 304
\bibitem[Henley \& Shelton (2008)]{hen08} Henley, D. B., \& Shelton, R. L. 2008, \apj, 676, 335
\bibitem[Hickox \& Markevitch (2006)]{hic06} Hickox, R. C., \& Markevitch, M. 2006, \apj, 645, 95
\bibitem[Hutchings \etal (2003)]{hut03} Hutchings, J. B., \etal 2003, \apj, 126, 2368
\bibitem[Ishisaki et al. (2007)]{Ishisaki_etal_2007} Ishisaki, K.  et al. 2007, \pasj, 59, S53
\bibitem[Jenkins (1978)]{jen78} Jenkins, E. B. 1978, \apj, 219, 845
\bibitem[Juett \etal (2006)]{juett2006} Juett, A., \etal 2006, \apj, 648, 1066
\bibitem[Koutroumpa \etal (2007)]{kou07} Koutroumpa, D., \etal 2007, A\&A, 475, 901
\bibitem[Koyama et al. (2007)]{Koyama_etal_2007} Koyama et al. 2007, \pasj, 59, S23
\bibitem[Kuntz \& Snowden (2001)]{kun01} Kuntz, K. D., \& Snowden, S. L. 2001, \apj, 554, 684
\bibitem[Lallement \etal (2004)]{lal04} Lallement, R., \etal 2004, A\&A, 426, 875
\bibitem[Leitherer \etal (1992)]{lei92} Leitherer, C., Bobert, C., \& Drissen, L 1992, \apj, 401, 596
\bibitem[Li \etal (2007)]{li07} Li, Z., Wang, Q. D., \& Hameed, S. 2007, \mnras, 376, 960
\bibitem[McCammon \etal (2002)]{mcc02} McCammon, D., \etal 2002, \apj, 576, 188
\bibitem[Mitsuda et al. (2006)]{mit06} Mitsuda, K.  et al. 2006, AAS, 208, 3910
\bibitem[Mitsuda et al. (2007)]{Mitsuda_etal_2007} Mitsuda, K.  et al. 2007, \pasj, 59, S1
\bibitem[Mitsuda et al. (2008)]{Mitsuda_etal_2008} Mitsuda, K.  et al. 2008, Prog. of  Theor.  Phys.  Suppl. , 169, 79
\bibitem[Mushotzky \etal (2000)]{mus00} Mushotzky, R. F., \etal 2000, Nature, 404, 459
\bibitem[Otte \& Dixon (2006)]{ott06} Otte, B., \& Dixon, W. V. D 2006, \apj, 647, 312
\bibitem[Page \etal (2003)]{page03} Page, M. J., \etal 2003, \mnras, 345, 639
\bibitem[Park \etal (1997)]{park97} Park, S., \etal 1997, \apjl, 476, L77
\bibitem[Rocks \etal (2007)]{roc07} Rocks, L. E., McCammon, D., Bauer, M., Fujimoto, R., \& Mitsuda, K. 2007, AAS, 20925418
\bibitem[Savage \etal (2003)]{sav03} Savage, B. D., \etal 2003, \apjs, 146, 125
\bibitem[Savage \etal (2005)]{sav05} Savage, B. D., \etal 2005, \apj, 619, 863
\bibitem[Savate \& Lehener (2006)]{sav06} Savage, B. D., \& Lehener, N. 2006, \apjs, 162, 134
\bibitem[Sembach \& Savage (1992)]{sem92} Sembach, K., \& B. D. Savage, 1992, \apjs, 83, 147
\bibitem[Shelton \etal (2001)]{she01} Shelton, R. L., \etal 2001, \apj, 560, 730
\bibitem[Shelton \etal (2007)]{she07} Shelton, R. L., Shallmen, S. M., \& Jenkins, E. B. 2007, \apj, 659, 365
\bibitem[Shelton (2008)]{she08} Shelton, R. L. 2008, SSRv, tmp, 69
\bibitem[Shull \& Slavin (1994)]{shu94} Shull, J. M., \& Slavin, J. D. 1994, \apj, 427, 784
\bibitem[Smith \etal (2007)]{smith07} Smith, R. K., \etal 2007, \pasj, 59, S141
\bibitem[Snowden \etal (1997)]{sno97} Snowden, S. L., \etal 1997, \apj, 485, 125
\bibitem[Sutherland \& Dopita (1993)]{sut93} Sutherland, R. S., \& Dopita, M. A. 1993, \apjs, 88, 253
\bibitem[Tawa et al. (2008)]{Tawa_etal_2008} Tawa et al.  2008, \pasj,  60, S11
\bibitem[T\"ullmann \etal (2006a)]{tul06a} T\"ullmann, R., \etal 2006a, A\&A, 448, 43
\bibitem[T\"ullmann \etal (2006b)]{tul06b} T\"ullmann, R., \etal 2006b, A\&A, 457, 779
\bibitem[Wang \etal (2001)]{wang01}  Wang, Q. D., \etal 2001, \apjl, 555, L99
\bibitem[Wang \etal (2003)]{wang02}  Wang, Q. D., \etal 2003, \apj, 598, 969
\bibitem[Wang \etal (2005)]{wang05}  Wang, Q. D., \etal 2005, \apj, 635, 386
\bibitem[Wilms \etal (2000)]{wil00} Wilms, J., Allen, A., \& McCray, R. 2000, \apj, 542, 914
\bibitem[Yao \& Wang (2005)]{yao05} Yao, Y., \& Wang, Q. D., 2005, \apj, 624, 751 
\bibitem[Yao \& Wang (2006)]{yao06} Yao, Y., \& Wang, Q. D., 2006, \apj, 641, 930 
\bibitem[Yao \& Wang (2007)]{yao07} Yao, Y., \& Wang, Q. D., 2007, \apj, 658, 1088 
\bibitem[Yao \etal (2008)]{yao08} Yao, Y., \etal 2008, \apjl, 672, L21
\end{thebibliography}
\end{document}